\documentclass[onecolumn,secnumarabic,amssymb, amsmath, nofootinbib,tightenlines,
nobibnotes, aps, prl,epsfig]{revtex4}
\usepackage{graphicx}
\usepackage{dcolumn}
\usepackage{bm}

\usepackage{amssymb}
\usepackage{epsfig}
\usepackage{color}

\newcommand{\ba}{\begin{eqnarray}}
\newcommand{\ea}{\end{eqnarray}}
\newcommand{\be}{\begin{equation}}
\newcommand{\ee}{\end{equation}}
\newcommand{\bdisplay}{\begin{displaymath}}
\newcommand{\edisplay}{\end{displaymath}}

\begin{document}
\preprint{APS/123-QED}
\title{Longitudinal Structure Function at the Limit $x=Q^2/s$}

\author{G. R. Boroun}%
 \email{ boroun@razi.ac.ir }
\affiliation{Department of physics, Razi University, Kermanshah
67149, Iran}

\date{\today}

 \pacs{***}
\begin{abstract}
The longitudinal structure function for nucleons and nuclei is
considered at fixed $\sqrt{s}$ and $Q^2$ to the minimum value of
$x$ given by $Q^2/s$. This is done using the expansion method and
color dipole model in the next-to-leading order approximation. The
extracted longitudinal structure functions were consistent with
HERA hepdata [ https://www.hepdata.net/record/ins1377206] and the
determination of $F_{L}$ at $x=Q^2/s$ [Frank E.Taylor, Phys. Rev.
D {\bf111}, 052001 (2025)] at moderate and large $Q^2$ values. The
results, consistent with the dipole picture at low $Q^2$ values,
show that the longitudinal structure function is small as expected
due to the transverse polarization of the exchanged photon and the
strong suppression of the dominant gluon component. Nonlinear
corrections to the nuclear longitudinal structure function at low
values of $x$ and $Q^2$ are also considered. These results may
enhance the deep inelastic scattering neutral current data in
future colliders at low $x$ and low $Q^2$. The longitudinal
structure functions for deuterium at low four-momentum transfer
squared, $Q^2<1~\mathrm{GeV}^2$ at the LO and NLO approximations
are determined and compared with the JLab
E00-002 data.\\

\end{abstract}
 \pacs{***}
\maketitle
\section{I. Introduction}

The longitudinal structure function, $F_{L}(x, Q^2)$, is a
fundamental observable in deep inelastic scattering (DIS)
experiments, as in the first approximation of the parton model, it
is equal identically zero \cite{Callan} since in the naive
quark-parton model (QPM) the massless spin-$\frac{1}{2}$ partons
cannot absorb the longitudinally polarized photon. The knowledge
of the longitudinal structure function at small values of the
Bjorken variable $x$ is important for understanding the inside
structure of hadrons (like protons) using high-energy lepton
collisions in future such as the Large Hadron electron Collider
(LHeC) \cite{LHeC} at CERN and the Electron-Ion Collider (EIC)
\cite{EIC} at Brookhaven National Laboratory (BNL) in e-A
scattering.\\
The longitudinal structure function helps test quantum
chromodynamics (QCD) predictions beyond leading order (LO) and
provides constraints on the gluon density crucial for LHC
predictions. It is sensitive to higher-twist effects and
non-linear QCD dynamics at very small $x$, and it is important for
precision Standard Model tests and searches for new physics. In
the QCD improved parton model, it provides important information
about the quarks and gluon distribution functions of the target.
The measured longitudinal structure function $F_{L}(x, Q^2)$ is
related to the cross section $\sigma_{L}$ for absorption of
longitudinally polarized virtual photon by
\ba \label{SigmaL_eq}
F_{L}(x,Q^2)=\frac{Q^2}{4{\pi^2}\alpha_{em}}(1-x)\sigma_{L}(x,Q^2),
\ea
and is related to the transverse structure functions by
\ba \label{Trans_eq} F_{L}(x,Q^2)=F_{2}(x,Q^2)-2xF_{1}(x,Q^2). \ea
The longitudinal structure function has been measured at HERA
(electron-proton collider) and in fixed-target experiments. It is
extracted from cross-section measurements at different beam
energies and determined in the high inelasticity ($y$) at the HERA
collider. HERA collected the electron-proton (ep) data at various
$x$ and $Q^2$ values for different $\sqrt{s}$ values, where $s$ is
the total energy squared of the electron-proton scattering. These
analyses have been conducted by the H1 \cite{H1} and ZEUS
\cite{ZEUS} collaborations. H1 covered a kinematic range from
$Q^2=1.5~\mathrm{GeV}^2$ and $x=0.279{\times}10^{-4}$ up to
$Q^2=800~\mathrm{GeV}^2$ and $x=0.0322$. ZEUS data has been taken
in much smaller region from $Q^2=9~\mathrm{GeV}^2$  up to
$Q^2=110~\mathrm{GeV}^2$. These kinematics will be extended down
to $x{\simeq}10^{-7}$ at the Future Circular Collider
electron-hadron (FCC-eh) \cite{FCC} with a center-of-mass energy
of $\sqrt{s}\simeq{3.5~\mathrm{TeV}}$ at a similar luminosity as
the LHeC with $\sqrt{s}\simeq{1.3~\mathrm{TeV}}$. This is about
four times the center-of-mass energy range of ep collisions at
HERA \cite{Klein, Armesto}. On the other hand, the center-of-mass
energy at the EIC will be approximately
$\sqrt{s}\simeq{140~\mathrm{GeV}}$ which is lower than the
flagship HERA data. These new colliders will enable the
 investigation of lepton-hadron processes
in ultra-high energy (UHE) neutrino astroparticle physics.\\
One of the main topics in hadron physics in the new accelerators
at small $x$ limit is the Color Glass Condensate (CGC) effective
field theory \cite{Iancu1, Salazar} in the dipole frame
\cite{Boris, Nikolaev}. Saturation models \cite{Bartels1, Munier}
and geometrical scaling \cite{Bartels2, Iancu2} can successfully
describe HERA data in the small $x$ and low $Q^2$ region. Future
electron-nucleus colliders are the best candidates for
discriminating between these models and the CGC physics. The CGC
forms the initial state, which is important in itself as a new
state of matter that depends on the unintegrated gluon
distribution (UGD). The next-to-leading-order (NLO)  corrections
to the cross section for the inclusive production of a pair of
hard jets using the color dipole picture for photon-nucleus
interactions at small $x$ are discussed in Ref.\cite{Iancu3}. The
importance of including a finite size for the target on
observables sensitive to small-$x$ evolution within the CGC is
discussed in Ref.\cite{Farid2}. The prospects for extracting the
longitudinal proton structure function  at the EIC, which is a
highly competitive direct probe of the proton gluon density, are
explored in Ref.\cite{Newman}.\\
The reduced cross section for inclusive e-p NC DIS (neutral
current deep inelastic scattering) depends on the DIS structure
functions, $F_{2}$ and $F_{L}$, as they include terms of photon,
photon-Z boson interference, and direct Z boson interaction terms.
In the kinematic range discussed above, the factors dependent on
Z-boson exchange are small. Therefore the reduced cross section is
as follows:
\ba \label{ReduCS_eq} \sigma_{r}(x,Q^2,s)=F_{2}(x,Q^
2)-\frac{y^2}{Y_{+}}F_{L}(x,Q^ 2), \ea
where  $Y_{+}=1+(1-y)^2$ and $y={Q^2}/{(xs)}$. Recently, Frank
E.Taylor \cite{Taylor} has discussed the longitudinal structure
function based on an extrapolation of the reduced neutral current
cross section with HERA data at a fixed $\sqrt{s}$ and $Q^2$ at
$x_{\mathrm{min}}=Q^2/s$, where indicate that the longitudinal
polarization of the virtual photon at $y=1$ is zero\footnote{The
virtual photon can be polarized into transverse and longitudinal
polarizations at low values of $x$ \cite{Hand}. The longitudinal
polarization of the virtual photon is given by
$\epsilon=\frac{2(1-y)}{Y_{+}}$.}. Therefore the measurement of
the reduced cross section in this limit is a determination of the
transversely polarized photon-parton scattering. The reduced cross
section, in this limit, is given by
\ba \label{ReduCSL_eq}
\mathrm{lim}_{y{\rightarrow}1}[\sigma_{r}(x,Q^2,s)]=F_{2}(Q^2/s,Q^
2)-F_{L}(Q^2/s,Q^ 2)=2\frac{Q^2}{s}F_{1}(Q^2/s,Q^ 2). \ea
This limit shows several properties consistent with the dipole
picture such as the effect of parton saturation \cite{Golec}  at
very low $Q^2$. A connection between the longitudinal DIS
structure function and the differential cross-section for
single-inclusive jet production with transverse momentum dependent
(TMD) is considered in Ref.\cite{Farid} at small Bjorken $x$. They
show that the single-inclusive jet cross-section in longitudinally
polarised DIS at leading power (LP) in $P_{\bot}/Q$ and small $x$
depends on gathering the quark and gluon jet contributions. Some
analytical solutions of the longitudinal DIS structure function
have been reported in recent years \cite{R1, R2, R3, R4, R5, R6,
R7, R8, R9, R10, R11, R12, R13, R14, R15}.\\
In DIS on nucleon targets, it is well known that an additional
constraint at $x_{\mathrm{min}}$ on the gluon distribution
determines the transversely polarized photon-parton scattering,
which comes from measuring the longitudinal structure function
$F_{L}$. On the other hand, due to the poor determination of the
nuclear gluon distribution, measurements of $F_{L}$ on nuclear
targets would be of great importance in EIC colliders. This would
help  constrain the gluon distribution and study the nuclear
dynamics at $x_{\mathrm{min}}$. In fact, studies within
perturbative QCD and model calculations show that the
corresponding nuclear effects closely follow those on the gluon
distribution at $x_{\mathrm{min}}$ \cite{Armesto}.\\
In this paper, an analytical distribution based on the expansion
method at the next-to-leading order (NLO) approximation is
extended to explore the longitudinal structure function at the
kinematic point $y=1$ (corresponding to $x_{\mathrm{min}}=Q^2/s$)
in nucleons and nuclei. The longitudinal structure function
approaches zero at this point, and this behavior could be
beneficial
for nucleon and nuclei parametrization groups.\\

\section{II. Method}

The longitudinal structure function can be related to the parton
distribution functions at small $x$ by ignoring the nonsinglet
distribution function. This relationship can be written as
\cite{AM, Gluck}
\ba \label{FL1_eq}
F_{L}(x,Q^2)=<e^2>\left[C_{L,q}(x,a_{s}){\otimes}x\Sigma(x,Q^2)+C_{L,g}(x,a_{s}){\otimes}xg(x,Q^2)\right],
\ea
where $a_{s}=\alpha_{s}/4\pi$ and
$<e^2>=\sum_{i=1}^{n_{f}}e_{i}^2/n_{f}$ with the charge $e_{i}$
for the active quark flavours ($n_{f}$). Here
$F_{2s}(x,Q^2)=x\Sigma(x,Q^2)=x\sum_{i=u,d,s,c,...}(q_{i}(x,Q^2)+\overline{q}_{i}(x,Q^2))$
and $G(x,Q^2)=xg(x,Q^2)$ are the singlet and gluon distribution
functions in the fractional hadron momentum. The perturbative
expansion of the coefficient functions is as follows
\ba \label{Coef_eq}
 C_{L,i}(x,a_{s})=\sum_{n=1}\left( \frac{\alpha_{s}}{4\pi}\right)^n
 c^{(n)}_{L,i}(x).
 \ea
where the coefficient functions in the LO approximation read
\cite{Moch}
 \ba
\label{CoefLO_eq}
C^{(1)}_{L,q}(x)=4C_{F}x,~~~~~~~~~~~~C^{(1)}_{L,g}(x)=8n_{f}x(1-x),
 \ea
where $C_{F}=4/3$ in QCD. The coefficient functions in the NLO
approximation at small fraction momentum read
  \ba
\label{CoefNLO_eq}
C^{(2)}_{L,q}(x)_{x{\ll}1}{\simeq}-2.370\frac{n_{f}}{x},~~~~~~~~~~~~C_{L,g}^{(2)}(x)_{x{\ll}1}{\simeq}-5.333\frac{n_{f}}{x}.
 \ea
By applying the convolution integral, the longitudinal structure
function can be rewritten as
  \ba
\label{FL2_eq}
F_{L}(x,Q^2)=\sum_{n=1}\left(
\frac{\alpha_{s}}{4\pi}\right)^n\int_{0}^{1-x{\approx}1}{dz}\left[
 c^{(n)}_{L,q}(1-z)F_{2}(\frac{x}{1-z},Q^2)+<e^2>c^{(n)}_{L,g}(1-z)G(\frac{x}{1-z},Q^2)\right].
 \ea
Expanding the distribution functions around the point $z=a$
\cite{Boroun1, Boroun2, Gay, Chen, BC, Boroun3} (where
$0{\leq}z{\lesssim}1$) can be done in the following form
\ba \label{Expandf_eq}
f(\frac{x}{1-z})|_{z=a}&=&f(\frac{x}{1-a})+\frac{x}{1-a}(z-a)\frac{{\partial}f(\frac{x}{1-a})}{{\partial}x}+\mathcal{O}(z-a)^{2}.
\ea
By integrating and then collecting terms, we ensure that the
following series converges for $|z-a|<1$ as
  \ba
\label{Convr_eq}
\frac{x}{1-z}|_{z=a}=\frac{x}{1-a}\sum_{k=1}^{\infty}\left[1+\frac{(z-a)^{k}}{(1-a)^{k}}\right].
\ea
Therefore, we find:
  \ba
\label{FL3_eq}
F_{L}(x,Q^2)&=&\frac{2\alpha_{s}}{3\pi}F_{2}\left(\frac{x}{1-a}(\frac{4}{3}-a),Q^2\right)+\frac{10\alpha_{s}}{27\pi}G\left(\frac{x}{1-a}(\frac{3}{2}-a),Q^2\right)\nonumber\\
&&-9.480\left(\frac{\alpha_{s}}{4\pi}\right)^2\frac{\frac{3}{2}-2a}{(1-a)^2}
F_{2}\left(\frac{\frac{3}{2}-2a}{(1-a)^2}x,Q^2\right)-5.926\left(\frac{\alpha_{s}}{4\pi}\right)^2\frac{\frac{3}{2}-2a}{(1-a)^2}
G\left(\frac{\frac{3}{2}-2a}{(1-a)^2}x,Q^2\right).
 \ea
At the point where $z=0.5$ is expanding (for furthers expanding
points refer to Appendix A), Eq.~(\ref{FL3_eq}) can be rewritten
in the following simplest form:
  \ba
\label{FL4_eq}
F_{L}(x,Q^2)&=&\frac{2\alpha_{s}}{3\pi}F_{2}(\frac{5}{3}x,Q^2)+\frac{10\alpha_{s}}{27\pi}G({2}x,Q^2)
-\frac{18.960\alpha^{2}_{s}}{16\pi^2}F_{2}(2x,Q^2)-\frac{11.852\alpha^{2}_{s}}{16\pi^2}G(2x,Q^2)\nonumber\\
&&=\frac{10\alpha_{s}}{27\pi}G({2}x,Q^2)\left[1+\frac{9}{5}\frac{F_{2}(\frac{5}{3}x,Q^2)}{G({2}x,Q^2)}\right]
-\frac{11.852\alpha^{2}_{s}}{16\pi^2}G(2x,Q^2)\left[1+\frac{18.960}{11.852}\frac{F_{2}(2x,Q^2)}{G(2x,Q^2)}\right]\nonumber\\
&&=\frac{10\alpha_{s}}{27\pi}G({2}x,Q^2)\left[1+\eta\right]
-\frac{11.852\alpha^{2}_{s}}{16\pi^2}G(2x,Q^2)\left[1+\xi\right].
\ea
At small $x$, the right-hand side is dominated by the gluon
contribution. In fact, the ratio of $F_{2}/G$ is less than $0.1$
at small $x$. The singlet and gluon distributions are related to
each other at small $x$ by the function $C(\alpha_{s}\ln{Q^2},
\alpha_{s},\alpha_{s}\ln{1/x})$ which describes the splitting of
the virtual photon into the quark-antiquark pair by the following
form \cite{Levin1, Levin2}:
  \ba
\label{Singlet_eq} F_{2s}(x,Q^2)&=&C(\alpha_{s}\ln{Q^2},
\alpha_{s},\alpha_{s}\ln{1/x} )xg(x,Q^2).
 \ea
Using the leading order DGLAP evolution equation that gives for
the proton structure function
  \ba
\label{SingF_eq}
 F_{2}(x,Q^2)&=&<e^2>\frac{C_{F}\alpha_{s}}{2\pi}\int_{0}^{\upsilon}d\upsilon'
 \int_{x}^{1}dz P_{qG}(z) \left(\frac{x}{z}g \left(
 \frac{x}{z},\upsilon'\right)\right),
 \ea
with $P_{qG}=(1+(1-z)^2)/z$ and $\upsilon={\ln}Q^2$. Here
$C_{F}=(N_{c}^{2}-1)/2N_{c}$ where $N_{c}$ is the number of
colors. At small values of $x$, the gluon distribution function
takes the form
  \ba
\label{GluonS_eq}
 xg(x,\upsilon)&=&\int_{\epsilon-i\infty}^{\epsilon+i\infty}\frac{d\gamma}{2\pi{i}}
  \left(\frac{1}{x}\right)^{\alpha_{s}\frac{N_{c}}{\pi}\chi(\gamma)}e^{\gamma{\upsilon}}g(\gamma)_{\mathrm{initial}},
 \ea
where the BFKL kernel $\chi(\gamma)$ has the following form
  \ba
\label{Chi_eq}
 \chi(\gamma)&=&2\psi(1)-\psi(\gamma)-\psi(1-\gamma),
 \ea
where for $\gamma{\rightarrow}\frac{1}{2}$
  \ba
\label{Chi_eq}
 \chi(\gamma)&{\rightarrow}&\omega_{0}+D\left(\gamma-\frac{1}{2}\right)^{2},
 \ea
with $\omega_{0}=4{\ln}2$ and $D=14\zeta(3)$. After integrating
and applying the diffusion approximation and the method of
steepest descent \cite{Levin2}, the simplest relation between the
proton structure function and the gluon distribution function is
obtained in the following form:
  \ba
\label{Chi_eq}
 \frac{F_{2}(x,Q^2)}{G(x,Q^2)}=<e^2>\frac{2C_{F}}{{\omega}_{0}N_{C}}{\simeq}0.09,
 \ea
with $n_{f}=4$. Instead of starting with a theoretically motivated
form of the color dipole model (CDM) in the dipole picture of deep
inelastic scattering \cite{Jeong}, we will begin with a
parametrization of the deep inelastic structure function for
electromagnetic scattering with protons due to the Bjorken
rescaling of the functions in Eqs.~(\ref{FL4_eq}, \ref{FLA0_eq},
\ref{FLA6_eq}). We can use the parametrizations of $F_{2}$ and $G$
by Donnachie-Landshoff (DL) \cite{DL} and Block-Durand-Ha (BDH)
\cite{Block}. The DL parametrization \cite{DL} meets the
requirement of being defined for all $Q^2$, even if not valid for
the full range, but it has a limited range of applicability. The
BDH parametrization applies to large and small $Q^2$ and small
$x$, where it has an expression for the asymptotic part of $F_{2}$
(no-valence) that accounts for the asymptotic behavior
($W^{2}{\rightarrow}\infty$ with $Q^2$ fixed). For small $x$,
where $W$ is the invariant mass of the final state,
$F_{2}(W^2,Q^2){\propto}\ln^{2}(W^2/Q^2){\simeq}\ln^{2}(1/x)$.
Taking the BDH parametrization, we find the coefficients
$\eta_{a}\left(\frac{Q^2}{s},Q^2 \right)$ and
$\xi_{a}\left(\frac{Q^2}{s},Q^2 \right)$ in Fig.1 over a wide
range of $Q^2$ at $x=Q^2/s$ with $\sqrt{s}=318~\mathrm{GeV}$. In
Fig.1, we observe that the ratio $\xi_{a}{\lesssim}0.1$ and
$\eta_{a}{\lesssim}0.12$ over a wide range of $Q^2$ values.
However, we can ignore the effects of these coefficients at
small values of $x$.\\
\begin{figure}
\centering
\includegraphics[width=0.55\textwidth]{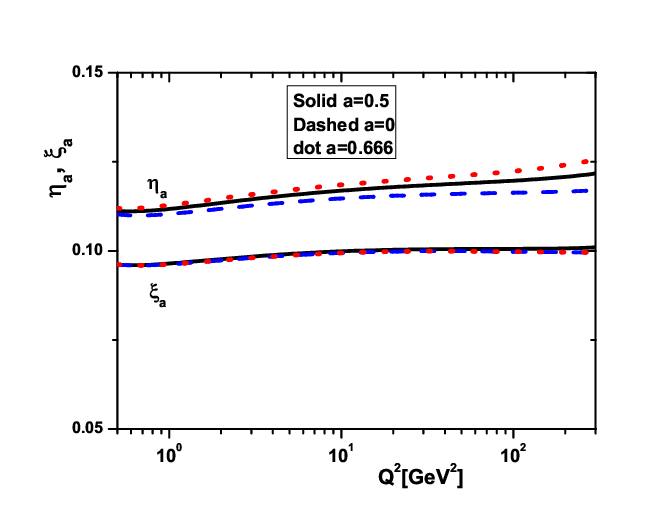}
\caption{Ratios of $\eta_{a}(\frac{Q^2}{s},Q^2)$ and
$\xi_{a}(\frac{Q^2}{s},Q^2)$ at $\sqrt{s}=318~\mathrm{GeV}$ as a
function of $Q^2$ with $a=0, 0.5$ and $0.666$ according to
Eqs.~(\ref{FLA0_eq}), (\ref{FL4_eq}) and (\ref{FLA6_eq}),
respectively .}\label{Fig1}
\end{figure}
Therefore, Eq.~(\ref{FL4_eq}) reduced and rewritten by the
following form
  \ba
\label{FL5_eq} F_{L}(Q^2/s,Q^2)&{\simeq}& \left[
\frac{10\alpha_{s}}{27\pi}-\frac{11.852\alpha^{2}_{s}}{16\pi^2}\right]G(2Q^2/s,Q^2),
\ea
where the gluon distribution is obtained using a dipole model fit
to low-$x$ data on $F_{2}$ in Ref.\cite{Thorn}.\\
To predict the rates of the various processes, a set of  gluon
distribution functions corresponding to the active flavor number
$n_{f}$ is required. When $Q^{2}$ increases above $m_{c}^{2}$ and
then above $m_{b}^{2}$, the number of active flavors increases
from $n_{f}=3$ to $n_{f}=4$ and then to $n_{f}=5$, which
corresponds to the variable-flavor-number scheme (VFNS)
\cite{Thorn2}. The charm and bottom quarks are considered as
infinitely massive below $Q^{2}=m^{2}_{c,b}$ and massless above
this threshold. For realistic kinematics it has to be extended to
the case of a general- mass VFNS (GM-VFNS) which is defined
similarly to the zero-mass VFNS (ZM-VFNS) in the
$Q^{2}/m^{2}_{c,b}{\rightarrow}\infty$ limit. For scales
$Q^{2}<m^{2}_{c}$ the fixed-flavor-number scheme (FFNS) is valid
and for $Q^{2}>m^{2}_{c}$, the approach outlined above to define a
VFNS is valid \cite{Martin}.\\
On the threshold of heavy quark production, the perturbative
predictions for longitudinal structure function are in accordance
with the heavy quark production threshold, which can be written as
\cite{R11, BFL}
  \ba
\label{FLc_eq} F_{L}(x,Q^{2})&=&C_{L,q}(x,a_{s})\otimes
F_{2}(x,Q^{2})+\frac{2}{9}C_{L,g}(x,a_{s})\otimes
G_{n_{f}=3}(x,Q^{2})+F_{L}^{c}(x,Q^{2}),~\mathrm{for}~Q^{2}{\geq}m_{c}^{2}
\ea
 and
  \ba
\label{FLcb_eq}
 F_{L}(x,Q^{2})&=&C_{L,q}(x,a_{s})\otimes
F_{2}(x,Q^{2})+\frac{2}{9}C_{L,g}(x,a_{s})\otimes
G_{n_{f}=3}(x,Q^{2})+F_{L}^{c}(x,Q^{2})+F_{L}^{b}(x,Q^{2}),~\mathrm{for}~Q^{2}{\geq}m_{b}^{2}.
\ea
 The HERA available data are insufficient to accurate extract
the $F_{L}^{c,b}$ values. Therefore, we include both charm and
bottom quark contributions to $F_{L}$, which are generated
radiatively from gluons in the color dipole model. The
relationship between the gluon distribution and the dipole
cross-section is defined by the dipole picture and the
$k_{T}$-factorization theorem \cite{Catani, Ellis}. The basic
theoretical formula for $F_{L}$ in terms of the longitudinal
photon polarization is given by
  \ba
\label{sigmaL_eq}
 F_{L}(x,Q^2)&{=}&
 \frac{Q^2}{4{\pi^2}\alpha_{\mathrm{em}}}\sigma_{L}^{\gamma^{*}p}(x,Q^2). \ea
When the virtual photon dissociates into a quark-antiquark pair (a
$q\overline{q}$ dipole), it subsequently interacts with the proton
as
  \ba
\label{sigmaL_eq}
 \sigma_{L}(x,Q^2)&{=}&\sum_{f}\int_{0}^{1}dz\int{d^{2}r}|\Psi_{L}(r,z,Q)|^{2}\widehat{\sigma}(x,r^2)
 , \ea
where the sum over quark flavors $f$ is performed as it depends on
the contributions from the light quark pairs ($u\overline{u},
d\overline{d}$ and $s\overline{s}$) as well as the contributions
from the $c\overline{c}$ and $b\overline{b}$ pairs. The photon
wave function depends on the mass of the quarks in the
$q\overline{q}$ dipole. Therefore, we can consider contributions
to the longitudinal structure function from the individual quark
flavor pairs according to Eqs.~(\ref{FLc_eq}) and (\ref{FLcb_eq})
where the Bjorken variable $x$ is modified to include the dipole
cross section into the quark mass using the following form
  \ba
\label{Bjorken_eq}
 x{\rightarrow}\widetilde{x}=x(1+\frac{4m_{f}^{2}}{Q^2}). \ea
Indeed, the Bjorken variable $x$ is modified at the limit
$x=Q^2/s$ to the rescaling variable $\widetilde{x}$, as
$\widetilde{x}=\frac{Q^{2}}{s}+\frac{4m_{f}^{2}}{s}$ which reduces
to the Bjorken variable $x$ at high $Q^2$ values. This
modification is consistent with the GM-VFNS in parameterization models.\\
The dipole cross-section is
  \ba
\label{sigmahat_eq}
 \widehat{\sigma}(x,r^2)&{=}&\frac{4\pi^2}{3}\int{\frac{dk^2}{k^4}}\alpha_{s}f(x,k^2)(1-J_{0}(kr))
 , \ea
where $J_{0}(kr)$ is the Bessel function of the first kind and
$f(x,k^2)$ is the unintegrated gluon distribution. The
unintegrated gluon distribution is obtained in the following form
\cite{Thorn}
  \ba
\label{UGD_eq} f(x,k^2)&=&\frac{3\sigma_{0}}{4{\pi^2}\alpha_{s}}
k^4(x/x_{0})^{\lambda}e^{-k^2(x/x_{0})^{\lambda}}. \ea
 The three parameters ($\sigma_{0}$, $x_{0}$ and $\lambda$) of the
fits with the GBW model using the quark masses are listed in Table
I.\\
\begin{table}
\centering \caption{The coefficient values are obtained in
Ref.\cite{Golec1}.}\label{table:table1}
\begin{minipage}{\linewidth}
\renewcommand{\thefootnote}{\thempfootnote}
\centering
\begin{tabular}{|l|c|c|c|c|c|c|c|} \hline\noalign{\smallskip}
fit &  $m_{l}[\mathrm{GeV}]$  &  $m_{c}[\mathrm{GeV}]$ & $m_{b}[\mathrm{GeV}]$  & $\sigma_{0}[\mathrm{mb}]$  & $\lambda$ &  $x_{0}/10^{-4}$  & $\chi^2/N_{\mathrm{dof}}$  \\
\hline\noalign{\smallskip}
0 &  0.14 &  - & - &  $23.58{\pm}0.28$ & $0.270{\pm}0.003$ & $2.24{\pm}0.16 $ & 1.83  \\
1 &  0.14 &  1.4 & - &  $27.32{\pm}0.35$ & $0.248{\pm}0.002$ & $0.42{\pm}0.04 $ & 1.60  \\
2 &  0.14 &  1.4 & 4.6 &  $27.43{\pm}0.35$ & $0.248{\pm}0.002$ & $0.40{\pm}0.04 $ & 1.61  \\
\hline\noalign{\smallskip}
\end{tabular}
\end{minipage}
\end{table}
The integrated gluon distribution is obtained using the leading
twist relation
  \ba
\label{gluonUGD_eq} xg(x,Q^2)&=&
\int_{0}^{Q^2}\frac{dk^2}{k^2}f(x,k^2). \ea
Therefore the integrated gluon distribution is obtained in the
following form
  \ba
\label{gluon_eq}
 G(x,Q^2)&{=}&
 \frac{3\sigma_{0}}{4{\pi^2}\alpha_{s}}\left[-Q^{2}e^{-Q^2(x/x_{0})^{\lambda}}+(x_{0}/x)^{\lambda}\left(1-e^{-Q^2(x/x_{0})^{\lambda}}
 \right)
\right]. \ea
The effect of charm and bottom quark contributions to the gluon
distribution are included by applying the rescaling variable as
  \ba
\label{gluonrescaling_eq} G({x},Q^2){\rightarrow}
G(\widetilde{x},Q^2)&{=}&
 \frac{3\sigma_{0}}{4{\pi^2}\alpha_{s}}\left[-Q^{2}e^{-Q^2(\widetilde{x}/x_{0})^{\lambda}}+(x_{0}/\widetilde{x})^{\lambda}\left(1-e^{-Q^2(\widetilde{x}/x_{0})^{\lambda}}
 \right)
\right]. \ea
Therefore, we find that  the longitudinal structure function at
the limit $x=Q^2/s$ depends on the mass of the quarks in the
$q\overline{q}$ dipole in the following form
\ba \label{FL6_eq}
 F_{L}(Q^2/s,Q^2)&{=}&
 \left[
\frac{10\alpha_{s}}{27\pi}-\frac{11.852\alpha^{2}_{s}}{16\pi^2}\right]\frac{3\sigma_{0}}{4{\pi^2}\alpha_{s}}[-Q^{2}e^{-Q^2(\frac{2}{x_{0}}(Q^2/s+4m_{f}^2/s))^{\lambda}}\nonumber\\
&&+(x_{0}/(2(Q^2/s+4m_{f}^2/s)))^{\lambda}\left(1-e^{-Q^2(\frac{2}{x_{0}}(Q^2/s+4m_{f}^2/s))^{\lambda}}\right)].
\ea

Recently, the authors in Ref.\cite{Peredo} modified the dipole
cross section with simulated nonlinear QCD evolution by
introducing a rescaling in the saturation scale as
$Q_{s}^{2}(x){\rightarrow}k \cdot Q_{s}^{2}(x)$ where
$Q_{s}^{2}(x)=Q_{0}^{2}(x/x_{0})^{-\lambda}$ with
$Q_{0}^{2}=1~\mathrm{GeV}^2$. The parameter $k$ can be understood
as a factor that controls the strength of the triple Pomeron
vertex and, therefore, the significance of nonlinear dynamics. A
value of $k=0$ corresponds to the linear behavior of the dipole
cross section, while $k=1$ results in the dipole cross sections
fitted to HERA data. Finally, for $k>1$,
 there is an additional enhancement of nonlinear effects.
 The nuclear enhancement of the saturation scale, which corresponds to an
 increase in the density of gluons, is defined on a large nucleus
rather than a single proton. In this case,
$Q_{s}^{2}(x){\rightarrow}Q_{s,A}^{2}(x)$, where $Q_{s,A}^{2}(x)$
represents the saturation scale for the nuclear target.\\
If the color dipole scatters on a large nucleus instead of a
single proton, then the saturation scales is modified by
\ba \label{Qsa_eq}
 Q^{2}_{s,A}(x)&{=}&k Q^{2}_{s}(x),
 \ea
where the parameter $k$ represents the enhancement of color
sources in the large nucleus for a nucleus with  mass number $A$,
where $k=A^{1/3}$. The study of the authors in Ref.\cite{Peredo}
is based on the Golec-Biernat Wusthoff (GBW) model \cite{Golec}
and Bartels Golec-Biernat Kowalski(BGK) model \cite{Bartels1}.
This allows us to more directly access the relevance of nonlinear
corrections for describing the energy dependence of the
photoproduction cross section. These models are modified by the
following forms:
\ba \label{SGBW_eq}
 \sigma^{\mathrm{GBW}}_{q\overline{q}}(x,r,k)&{=}&\frac{\sigma^{\mathrm{GBW}}_{0}}{k}
 \left[1-\exp\left(-k.\frac{r^2Q_{s}^{2}(x)}{4} \right)\right],
\ea
and
\ba \label{SBGK_eq}
 \sigma^{\mathrm{BGK}}_{q\overline{q}}(x,r,k)&{=}&\frac{\sigma^{\mathrm{BGK}}_{0}}{k}
 \left[1-\exp\left(-k.\frac{r^2{\pi^2}\alpha_{s}(\mu^{2}_{r})xg(x,\mu^2_{r})}{3\sigma^{\mathrm{BGK}}_{0}} \right)\right],
\ea
where the renormalization scale $\mu$ is usually identified with
the factorization scale and taken to depend on the dipole size
with $\mu^2{\sim}1/r^2$ for small dipole sizes. Now we use the DIS
$F_{L}$ on electron-proton collisions to make predictions for
electron-ion collisions with the replacements \cite{Carvalho1}
$S{\rightarrow}S_{A}=A^{2/3}S$, where $S$ is the area of the
target and the transverse size of the dipole cross section scaling
is approximately proportional to $\sim{A^{2/3}}$, and
$Q_{s}^{2}(x){\rightarrow}Q_{s,A}^{2}(x)=A^{1/3}Q_{s}^{2}(x)$. The
nuclear longitudinal structure function at the limit $x=Q^2/s$ is
\ba \label{FLNUC_eq}
 F^{A}_{L}(Q^2/s,Q^2)&{=}&
 \left[
\frac{10\alpha_{s}}{27\pi}-\frac{11.852\alpha^{2}_{s}}{16\pi^2}\right]\frac{3\sigma_{0}A^{2/3}}{4{\pi^2}\alpha_{s}}[-Q^{2}e^{-\frac{Q^2}{A^{1/3}}
(\frac{2}{x_{0}}(Q^2/s+4m_{f}^2/s))^{\lambda}}\nonumber\\
&&+A^{1/3}(x_{0}/(2(Q^2/s+4m_{f}^2/s)))^{\lambda}\left(1-e^{-\frac{Q^2}{A^{1/3}}(\frac{2}{x_{0}}(Q^2/s+4m_{f}^2/s))^{\lambda}}\right)].
\ea
The nuclear DIS structure functions in recent years by several
methods have been proposed in Refs.\cite{Armesto1, Armesto2,
Rezaei, nCTEQ, Khanpour, Armesto3, Machado, Nikolaev2}. In
Ref.\cite{Armesto1}, the longitudinal structure function in
nuclear DIS at small $x$ is discussed within the framework of
universal parton densities obtained in DGLAP analysis. The
longitudinal structure function in pQCD in the
$\overline{\mathrm{MS}}$ scheme was calculated, and the ratio
$F_{L}^{A}/AF_{L}^{p}$ was compared with PDF sets such as EPS09
\cite{EPS}, nDS \cite{DS}, HKN07 \cite{HKN}, and FGS10 \cite{FGS}
for lead (A=208). In Ref.\cite{Farid}, the authors studied of TMD
factorization in the target fragmentation region and found an
interesting result concerning the leading power contribution to
the unintegrated $F_{L}$ distribution at NLO. The NLO computations
of the unintegrated $F_{L}$ at leading power are consistent with
the Altarelli-Martinelli relation at NLO which included a step in
DGLAP for the gluon distribution at LO and the contribution from
sea quarks coming from the splitting of gluon into a
quark-antiquark pair. In Ref.\cite{Machado2}, the author found the
longitudinal structure function at small $x$ by the gluon content
of the nucleon target at fixed energy within the color dipole
formalism. In the following we extend these results to the nuclear
longitudinal structure function for comparison.\\
In Ref.\cite{ECA}, the authors simulated the nuclear structure
functions based on measurements of inclusive and charm reduced
cross sections at an EIC. Since $F_{L}$ has a larger direct
contribution from gluons, information on $F_{L}$ and,
consequently, direct access to nuclear gluons are not currently
available. However, at an EIC, the high luminosity and wide
kinematic reach will enable the direct extraction of $F_{L}$,
providing more information on the behavior of nuclear gluons. The
potential for extracting the longitudinal proton structure
function at EIC through a Rosenbluth separation method was
explored in Ref.\cite{Newman}. Simulated extractions of the
longitudinal structure function for three different pseudo-data
replicas in the conservative scenario are presented, with
different colors corresponding to different replicas.\\

\section{III. Results and Discussion}

The standard representation for QCD couplings in the LO and NLO
(within the $\overline{\mathrm{MS}}$-scheme) approximations reads
\begin{eqnarray}
\alpha_{s}(t)=\frac{4\pi}{\beta_{0}t}~~~~~~~~~~~~~~~~~~~~~~~~~~(\mathrm{LO}),\nonumber\\
\alpha_{s}(t)=\frac{4\pi}{\beta_{0}t}\bigg{[}1-\frac{\beta_{1}
{\ln}{\ln}(t)}{\beta_{0}^{2}t}\bigg{]}~~~~~~(\mathrm{NLO}),\nonumber
\end{eqnarray}
with $\beta_{0}$ and $\beta_{1}$ as the first two coefficients of
the QCD $\beta$-function,
\begin{eqnarray}
\beta_{0}=\frac{1}{3}(11C_{A}-n_{f}),
~~~\beta_{1}=\frac{1}{3}(34C_{A}^{2}-2n_{f}(5C_{A}+3C_{F})),\nonumber
\end{eqnarray}
where $C_{F}=\frac{N_{c}^{2}-1}{2N_{c}}$ and $C_{A}=N_{c}$ are the
Casimir operators in the fundamental and adjoint representations
of the $\mathrm{SU(N_{c})}$ color group, and
$t=\ln\frac{Q^{2}}{\Lambda^{2}}$ where the QCD parameter $\Lambda$
for active flavors has been determined \cite{R3} based on
$\alpha_{s}(M_{z}^{2})=0.1166$,
\begin{eqnarray}
\mathrm{LO} :
\Lambda(n_{f}=3)=0.1368~\mathrm{GeV},~\Lambda(n_{f}=4)=0.1368~\mathrm{GeV},
\Lambda(n_{f}=5)=0.8080~\mathrm{GeV},\nonumber\\
\mathrm{NLO} :
\Lambda(n_{f}=3)=0.3472~\mathrm{GeV},~\Lambda(n_{f}=4)=0.2840~\mathrm{GeV},
\Lambda(n_{f}=5)=0.1957~\mathrm{GeV}.
\end{eqnarray}
\begin{figure}
\centering
\includegraphics[width=0.55\textwidth]{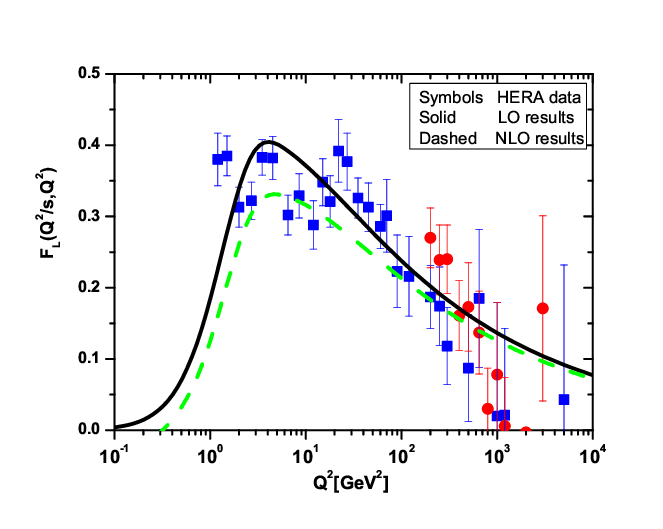}
\caption{The DIS longitudinal structure functions at LO (black
solid curve) and NLO (green dashed curve) approximations are shown
as a function of $Q^2$ for $\sqrt{s}=318~\mathrm{GeV}$. These
results consider the charm effect in the rescaling of the Bjorken
 variable $x$. The coefficients
  are based on the results from Fit 1 in Table I. The HERA combined H1 and ZEUS reduced cross section datasets (squares-blue and circles-red) and their uncertainties
are shown.}\label{Fig2}
\end{figure}
\begin{figure}
\centering
\includegraphics[width=0.55\textwidth]{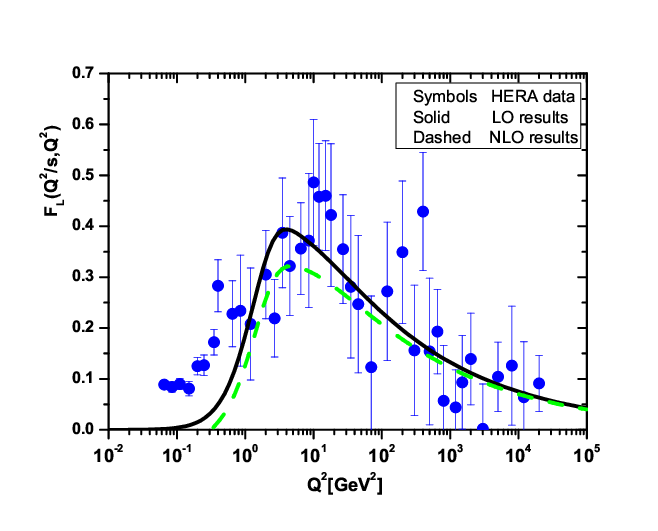}
\caption{The same as Fig.1 at
$\sqrt{s}=300~\mathrm{GeV}$.}\label{Fig3}
\end{figure}
In Figs.2 and 3, we compare the longitudinal structure function at
the limit of $x=Q^2/s$ for $\sqrt{s}=318$ and $300~\mathrm{GeV}$
respectively. This comparison is made with the $e^{+}p$ HERA data,
as shown in Tables 1 and 2 of Ref.\cite{HERA}. These tables
combine the reduced cross section datasets from H1 and ZEUS, along
with their respective kinematic regions. The extracted
longitudinal structure functions are shown as a function of $Q^2$
and $x_{\mathrm{min}}$ in Tables VIII and IX of Ref.\cite{Taylor},
accompanied by the uncertainties of the data points. The results
at the LO and NLO approximations are obtained from the gluon
distribution behavior using a dipole picture \cite{Thorn}. We
observe that this distribution accurately represents the
longitudinal structure function at both low and high $Q^2$ values,
consistent with the HERA data. The longitudinal structure
functions obtained from the dipole picture show that $F_{L}$ is
small at low $x$ and low $Q^2$ because the polarization of the
exchanged photon is transverse at this kinematic point. The
results in Figs.2 and 3 correspond to the four active flavors
(i.e., $n_{f}=4$) while considering the rescaling variable, i.e.,
Eq.~(\ref{Bjorken_eq}).\\
\begin{figure}
\centering
\includegraphics[width=0.55\textwidth]{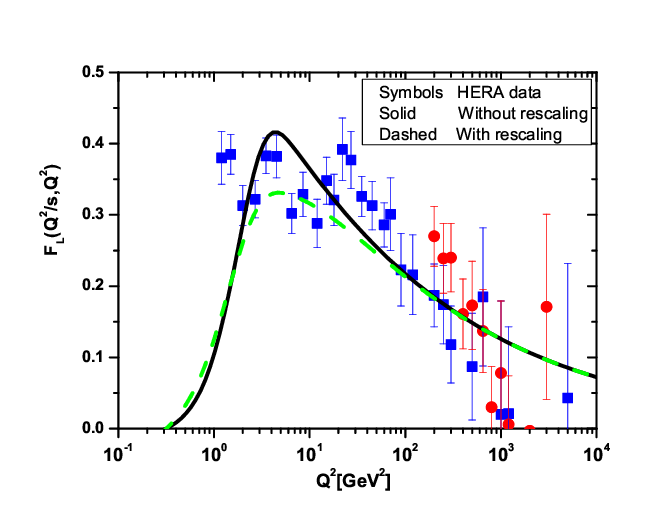}
\caption{The DIS longitudinal structure functions without
rescaling (black solid curve) and with rescaling (green dashed
curve) approximations at the NLO approximation are shown as a
function of $Q^2$ for $\sqrt{s}=318~\mathrm{GeV}$. The
coefficients
  are based on the results from Fit 1 in Table I. The HERA combined H1 and ZEUS reduced cross section datasets (squares-blue and circles-red) and their uncertainties
are shown.}\label{Fig4}
\end{figure}
The effects of the rescaling variable in the longitudinal
structure function are shown in Fig.4. We observe that the
behavior of  $F_{L}(Q^2/s,Q^2)$ is affected by the large quark
mass $m_{c}{\approx}1.4~\mathrm{GeV}$, whether or not the
rescaling variable is used. When comparing with HERA data, the
comparison  without rescaling shows lower quality at moderate
$Q^2$. However, the comparison is significantly improved with
scaling.\\
This rescaling strongly depends on the mass of the quarks in the
$q\overline{q}$ dipole. In Fig.5, we show the mass effects of the
rescaling variable in  $F_{L}(Q^2/s,Q^2)$. We observe that adding
charm into the analysis improves the results at low and moderate
$Q^2$ values compared  to  the contributions of light and bottom
quarks. Nevertheless, the parameters of the bottom quark
contribution can be taken into account for further analysis in
order to have a full understanding of heavy quarks
\cite{Golec1}.\\
\begin{figure}
\centering
\includegraphics[width=0.55\textwidth]{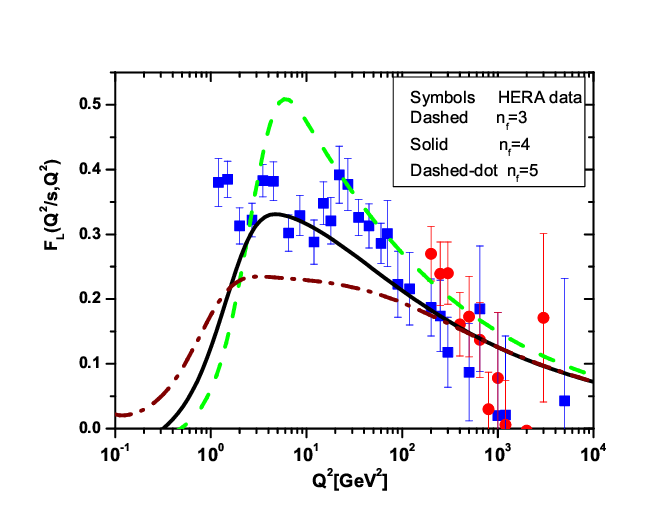}
\caption{The DIS longitudinal structure functions at the NLO
approximation are shown by applying the active flavors ($n_{f}=3$,
green dashed curve), ($n_{f}=4$, black solid curve) and
($n_{f}=5$, brown dashed-dot curve) as a function of $Q^2$ for
$\sqrt{s}=318~\mathrm{GeV}$ based on the results  in Table I and
QCD cut-off. The HERA combined H1 and ZEUS reduced cross section
datasets (squares-blue and circles-red) and their uncertainties
are shown.}\label{Fig5}
\end{figure}
\begin{figure}
\centering
\includegraphics[width=0.55\textwidth]{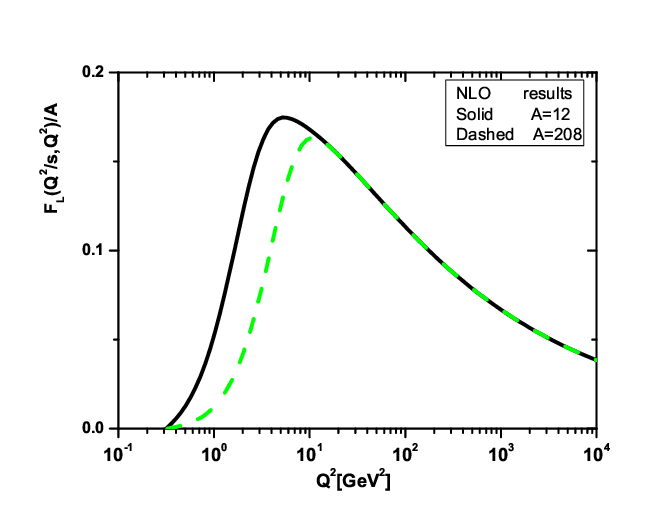}
\caption{The DIS longitudinal structure functions for light nuclei
C-12 (black solid curve) and heavy nuclei Pb-208 (green dashed
curve) at the  NLO  approximation are shown as a function of $Q^2$
for $\sqrt{s}=89~\mathrm{GeV}$. In these results the charm effect
in the rescaling Bjorken
 variable $x$ is considered. The coefficients
  are based on the results from Fit 1 in Table I.}\label{Fig6}
\end{figure}
In Fig.6, we show the behavior of the DIS longitudinal structure
function divided by A at the NLO approximation for light nuclei
C-12 and heavy nuclei Pb-208 at $x_{\mathrm{min}}$ in the
kinematic range relevant for the EIC
($\sqrt{s}=89~\mathrm{GeV}$)\footnote{The center-of-mass energies
in electron-ion colliders proposed in China and the US are $15
-20~\mathrm{GeV}$ for EIcC and $30 -140~\mathrm{GeV}$ for EIC
respectively.}, taking into account the rescaling variable due to
the charm quark mass. We observe that the behavior of the nuclear
longitudinal structure function, $F^{A}_{L}(Q^2/s,Q^2)/A$,
determines the transversely polarized photon-nuclei scattering at
low $Q^2$ values. The enhancement of $F^{A}_{L}(Q^2/s,Q^2)/A$
decreases at moderate $Q^2$ values as the mass number A increases,
and increases as $Q^2$ values increase. The behavior of
$F^{A}_{L}(Q^2/s,Q^2)/A$ is independent of the mass number A  at
large $Q^2$ values.\\
\begin{figure}
\centering
\includegraphics[width=0.55\textwidth]{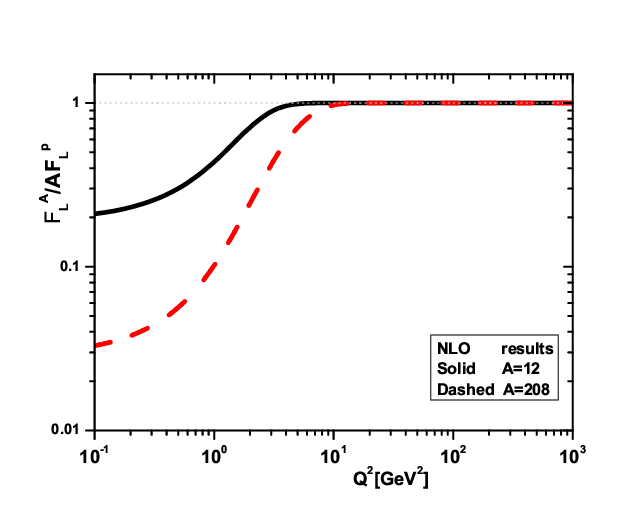}
\caption{ The ratio of
$\frac{F^{A}_{L}}{AF^{p}_{L}}(\frac{Q^2}{s}+\frac{4m_{c}^{2}}{s},Q^2)$
for A=12 (black solid curve) and A=208 (red dashed curve) is shown
as a function of $Q^2$ for $\sqrt{s}=89~\mathrm{GeV}$ and compared
with the gluonic term of the CS results \cite{Farid} which is
modified for A=12 (blue dashed-dot curve) and A=208 (brown short
dashed curve). }\label{Fig7}
\end{figure}
Indeed, the parameter $k=A^{1/3}$ in Eq.~(\ref{FLNUC_eq})
implements the enhancement of the saturation scale based on the
gluon saturation, corresponding to an increase in the density of
gluons \cite{Peredo}. This saturation is visible in large nuclei
at low $Q^2$ values, as shown in Fig.6. The gluon saturation is
tested by comparing the longitudinal structure functions in
protons and nuclei. In Fig.7, the behavior of the ratio
$\frac{F^{A}_{L}}{AF^{p}_{L}}$ is considered at the minimum value
of $x$ given by $Q^2/s$ at the EIC center-of-mass energy
$\sqrt{s}=89~\mathrm{GeV}$. A clear deviation of the ratio
$\frac{F^{A}_{L}}{AF^{p}_{L}}$  from the linearized description,
which indicates the relevance of nonlinear terms, is visible in
Fig.7 and is significant for large nuclei. The nonlinear effects
begin for light and heavy nuclei at $Q^2<10~\mathrm{GeV}^2$. The
deviation of this ratio from unity demonstrates the significance
of nonlinear effects,  indicating  saturation physics at
$x_{\mathrm{min}}$. The results of saturation effects in eA
collisions, conducted using the color dipole approach, confirmed
the findings presented in Ref.\cite{Carvalho2}. The effect of
different values of $k$ ($k=2.289$ and $k=5.925$) on the behavior
of the ratio at $x_{\mathrm{min}}$ is visible in Fig.7 and may
already be observable for photonuclear reactions, where gluon
densities are enhanced by $A^{1/3}$. These results are compared
with the gluonic term of the P.Caucal and F.Salazar (CS) results
\cite{Farid} that is modified for the nuclear longitudinal
structure function. This modification is based on the TMD
factorization in the target fragmentation region
and consistency with the Altarelli-Martinelli relation at NLO.\\
Measurements of the longitudinal structure function for the
deuterium at low values of $Q^2$ and intermediate $x$ were
presented in Ref.\cite{JLAB} where the experiment was conducted at
Jefferson Lab (JLab). In this experiment, the separated structure
functions for hydrogen and deuterium using the Rosebluth
separation technique at low four-momentum transfer squared,
$Q^2<1~\mathrm{GeV}^2$, were analyzed and compared with parton
distribution parameterizations and a $k_{T}$ factorization
approach. In Fig.8, we compare our results with JLab data for the
longitudinal structure function of deuterium as a function of $x$
(for $x>0.01$) for two values of $Q^2=0.4$ and
$0.5~\mathrm{GeV}^2$. The longitudinal structure function for
nuclei presented in this work (i.e., Eq.~(\ref{FLNUC_eq})) is
based on the color dipole scattering on a large nucleus
\cite{Peredo} where the saturation scale increases due to the
nuclear "oomph factor". Therefore, when compared to the deuterium
longitudinal structure function at low $Q^2$ values, we expect to
see differences in the results. In Fig.8, we observe that the JLab
E00-002 data are plotted between the LO and NLO results.\\
To further investigate the behavior of the results with $Q^2$ and
compare them with JLab data, we present the longitudinal structure
function results for deuterium versus $Q^2$ for $x=0.1$ and $0.4$
at the NLO approximation in Fig.9. The data in Fig.9 were obtained
from the averaged JLab experiment ( E00-002 data). In this
kinematic regime, the longitudinal structure function
$F_{L}^{D}(x,Q^2)$ is well described by the JLab data. We observe
that, due to current conservation, the interaction of longitudinal
virtual photons vanishes in the limit of real photon $Q^2=0$.\\
\begin{figure}
\centering
\includegraphics[width=0.55\textwidth]{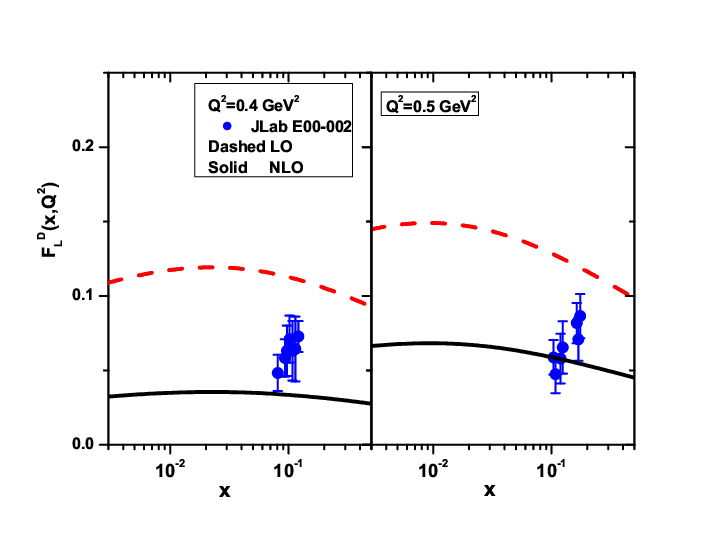}
\caption{ Longitudinal structure function results for deuterium
are shown for fixed $Q^2$ as a function of Bjorken $x$ at the LO
(dashed curve) and NLO (solid curve) approximations. The two
panels correspond to two different $Q^2$ regions:
$Q^2=0.4~\mathrm{GeV}^2$ (left) and $Q^2=0.5~\mathrm{GeV}^2$
(right). In this kinematic regime, the data are from Jefferson Lab
\cite{JLAB} and are accompanied by total errors. }\label{Fig8}
\end{figure}
\begin{figure}
\centering
\includegraphics[width=0.65\textwidth]{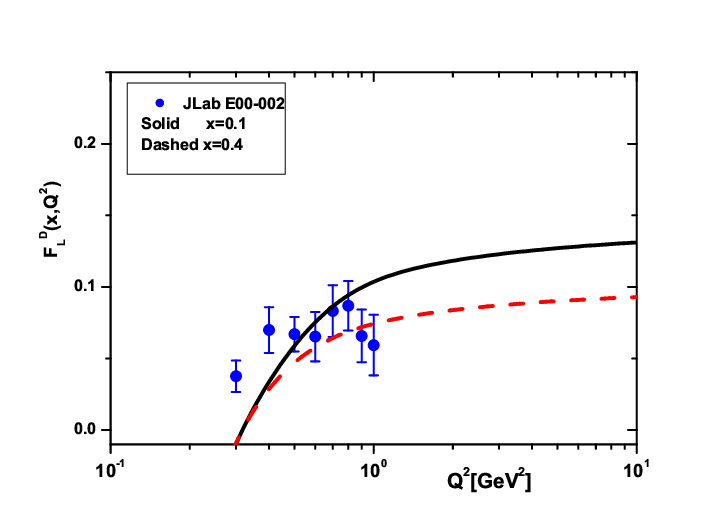}
\caption{ The results for the deuterium longitudinal structure
function are displayed for a fixed $x$ as a function of $Q^2$ at
the NLO approximation. Two curves are presented, representing two
distinct $x$ values: $x=0.1$ (black solid curve) and $x=0.4$ (red
dashed curve). The averaged data  for $x{\geq}0.05$ are sourced
from Jefferson Lab \cite{JLAB} and include total errors.
}\label{Fig9}
\end{figure}

In conclusion, we have examined the behavior of the longitudinal
structure function of protons and nuclei at $y=1$, where
$x_{\mathrm{Bj}}=x_{\mathrm{min}}=Q^2/s$, in both HERA and EIC
kinematics. We utilized the expanding method and CDP  for low $x$
gluon distributions in this scenario. Our model provides a good
description of results at moderate and large $Q^2$ values when
compared to HERA data, and predicts nuclear longitudinal structure
functions that can be measured in electron-ion collisions at
$x_{\mathrm{min}}$. Comparing with HERA data at moderate and large
$Q^2$ values strongly suggests that the gluon component is the
dominant factor in the longitudinal structure function within the
dipole picture. We observed that the value of $F_{L}$ is small at
low $Q^2$ values in the dipole picture. This is because the
dominant gluon component is strongly suppressed and the
polarization of the exchanged photon is transverse at this
kinematic point. We investigated the nuclear longitudinal
structure function at this limit, which reveals high gluon
densities and associated nonlinear high-energy evolution. The
nonlinear corrections increase with the mass number $A$. The ratio
$\frac{F^{A}_{L}}{AF^{p}_{L}}$ can serve as an indicator for the
presence of nonlinear low $x$ dynamics in large nuclei. The
longitudinal structure functions for deuterium at low
four-momentum transfer squared in the LO and NLO approximations
are determined and compared with
the JLab E00-002 data.\\

\section{Appendix A}
The longitudinal structure function at the expanding points $z=0$
and $0.666$ is respectively:
  \ba
\label{FLA0_eq}
F_{L}(x,Q^2)&=&\frac{2\alpha_{s}}{3\pi}F_{2}(\frac{4}{3}x,Q^2)+\frac{10\alpha_{s}}{27\pi}G(\frac{3}{2}x,Q^2)
-\frac{28.440\alpha^{2}_{s}}{32\pi^2}F_{2}(\frac{3}{2}x,Q^2)-\frac{17.778\alpha^{2}_{s}}{32\pi^2}G(\frac{3}{2}x,Q^2)\nonumber\\
&&=\frac{10\alpha_{s}}{27\pi}G(\frac{3}{2}x,Q^2)\left[1+\frac{9}{5}\frac{F_{2}(\frac{4}{3}x,Q^2)}{G(\frac{3}{2}x,Q^2)}\right]-
\frac{17.778\alpha^{2}_{s}}{32\pi^2}G(\frac{3}{2}x,Q^2)\left[1+\frac{28.440}{17.778}\frac{F_{2}(\frac{3}{2}x,Q^2)}{G(\frac{3}{2}x,Q^2)}\right]\nonumber\\
&&=\frac{10\alpha_{s}}{27\pi}G(\frac{3}{2}x,Q^2)\left[1+\eta\right]-
\frac{17.778\alpha^{2}_{s}}{32\pi^2}G(\frac{3}{2}x,Q^2)\left[1+\xi\right],
\ea
and
  \ba
\label{FLA6_eq}
F_{L}(x,Q^2)&=&\frac{2\alpha_{s}}{3\pi}F_{2}(\frac{4}{2}x,Q^2)+\frac{10\alpha_{s}}{27\pi}G(\frac{5}{2}x,Q^2)
-\frac{14.315\alpha^{2}_{s}}{16\pi^2}F_{2}(\frac{3}{2}x,Q^2)-\frac{8.948\alpha^{2}_{s}}{16\pi^2}G(\frac{3}{2}x,Q^2)\nonumber\\
&&=\frac{10\alpha_{s}}{27\pi}G(\frac{5}{2}x,Q^2)\left[1+\frac{9}{5}\frac{F_{2}(\frac{4}{2}x,Q^2)}{G(\frac{5}{2}x,Q^2)}\right]-
\frac{8.948\alpha^{2}_{s}}{16\pi^2}G(\frac{3}{2}x,Q^2)\left[1+\frac{14.315}{8.948}\frac{F_{2}(\frac{3}{2}x,Q^2)}{G(\frac{3}{2}x,Q^2)}\right]\nonumber\\
&&=\frac{10\alpha_{s}}{27\pi}G(\frac{5}{2}x,Q^2)\left[1+\eta\right]-
\frac{8.948\alpha^{2}_{s}}{16\pi^2}G(\frac{3}{2}x,Q^2)\left[1+\xi\right].
\ea
 %


\subsection{ACKNOWLEDGMENTS}

The author would like to thank  F.E.Taylor and F.Salazar for their
helpful comments and invaluable support. The author is especially
grateful to Nestor Armesto, Elke Aschenauer, and Paul Newman for
 fruitful discussions and for allowing access to data
related to the longitudinal structure function at the EIC and
JLab. I am also very grateful to the Department of Physics at
CERN-TH for their
warm hospitality.\\

\end{document}